\documentclass[prl,twocolumn,superscriptaddress,showpacs,floatfix,longbibliography]{revtex4-1}
\usepackage{mathrsfs,braket}
\usepackage{amssymb, amsbsy, amsmath, latexsym, dsfont, array, layout,
graphicx,mathrsfs,color,ulem,bm}
\usepackage[colorlinks=true,citecolor=blue,urlcolor=blue]{hyperref}

\begin{document}

\title{Simulating non-Hermitian quasicrystals with photonic quantum walks}

\author{Quan Lin}
\affiliation{Beijing Computational Science Research Center, Beijing 100084, China}
\author{Tianyu Li}
\affiliation{CAS Key Laboratory of Quantum Information, University of Science and Technology of China, Hefei 230026, China}
\affiliation{CAS Center For Excellence in Quantum Information and Quantum Physics, Hefei 230026, China}
\author{Lei Xiao}
\affiliation{Beijing Computational Science Research Center, Beijing 100084, China}
\author{Kunkun Wang}
\affiliation{Beijing Computational Science Research Center, Beijing 100084, China}
\author{Wei Yi}\email{wyiz@ustc.edu.cn}
\affiliation{CAS Key Laboratory of Quantum Information, University of Science and Technology of China, Hefei 230026, China}
\affiliation{CAS Center For Excellence in Quantum Information and Quantum Physics, Hefei 230026, China}
\author{Peng Xue}\email{gnep.eux@gmail.com}
\affiliation{Beijing Computational Science Research Center, Beijing 100084, China}

\begin{abstract}
Non-Hermiticity significantly enriches the properties of topological models, leading to exotic features such as the non-Hermitian skin effects and non-Bloch bulk-boundary correspondence that have no counterparts in Hermitian settings. Its impact is particularly illustrating in non-Hermitian quasicrystals where the interplay between non-Hermiticity and quasiperiodicity results in the concurrence of the delocalization-localization transition, the parity-time (PT)-symmetry breaking, and the onset of the non-Hermitian skin effects. Here we experimentally simulate non-Hermitian quasicrystals using photonic quantum walks. Using dynamic observables, we demonstrate that the system can transit from a delocalized, PT-symmetry broken phase that features non-Hermitian skin effects, to a localized, PT-symmetry unbroken phase with no non-Hermitian skin effects. The measured critical point is consistent with the theoretical prediction through a spectral winding number, confirming the topological origin of the phase transition. Our work opens the avenue of investigating the interplay of non-Hermiticity, quasiperiodicity, and spectral topology in open quantum systems.
\end{abstract}

\maketitle

Non-Hermiticity arises in open systems and can lead to intriguing properties with no Hermitian counterparts~\cite{QJ,benderreview,bender,photonpt1,review2,review3,nonHtopo1,nonHtopo2,nonHtopo3, WZ1,WZ2,Budich,mcdonald,alvarez,murakami,ThomalePRB, fangchenskin,kawabataskin,Slager,yzsgbz,stefano,tianshu,lli,luole,yuce,ngra,nsa,nta}. Exotic non-Hermitian phenomena such as the parity-time (PT) symmetry and exceptional points~\cite{bender,luole,yuce,photonpt1,review3,nonHtopo1,nonHtopo2,nonHtopo3}, the non-Hermitian skin effects~\cite{WZ1,WZ2,Budich,mcdonald,alvarez,murakami,ThomalePRB,fangchenskin,kawabataskin,Slager,yzsgbz,stefano,tianshu,lli}, and the non-Bloch bulk-boundary correspondence~\cite{WZ1,WZ2} have attracted much attention. The interest is further stimulated by their experimental observation in open systems~\cite{teskin,teskin2d,metaskin,photonskin,XDW+21,scienceskin,nsm}, with potential applications in precision measurements, non-reciprocal quantum device, and topological transport. So far, these experiments either focus on non-Hermitian models with no spatial degrees of freedom, or on lattices with discrete translational symmetry. The properties of non-Hermitian quasicrystals remain largely unexplored experimentally.

In closed quantum systems, quasicrystals play host to a wealth of properties~\cite{harper,AA,sokoloff,roati,lahini,verbin,nsl,naad}. For instance, in the one-dimensional Aubry-Andr\'e-Harper (AAH) model, a finite strength of quasiperiodicity drives the system from an extended metallic state into an Anderson insulator where the bulk eigenstates become exponentially localized~\cite{harper,AA,sokoloff,nsl,naad}. The AAH model can also be mapped to the Hofstadter model, and hence to the integer quantum Hall system, revealing a deeper connection with band topology~\cite{hofstadter,kraus,chen1}. Variations of the AAH model feature different forms of quasiperiodic disorder, as well as interactions, which can further give rise to more exotic phases such as the many-body localized~\cite{mbl1,mbl2,mbl3} or the critically-localized states~\cite{crit1,crit2,crit3}. In a recent series of theoretical studies, it has been shown that specific non-Hermitian variants of the AAH model can feature a topological phase transition characterized by a spectral winding number~\cite{nonHtopo1,chen3,chen4,chen5,YX,stefano2}. Remarkably, such a critical point is a simultaneous demarcation for the delocalization-localization transition, the PT-symmetry-breaking transition, as well as for the onset of the non-Hermitian skin effects when an open boundary condition (OBC) is imposed~\cite{chen3}. Given the uniqueness of such a multi-critical point, the experimental simulation of non-Hermitian quasicrystals and the underlying rich critical physics is desirable.

In this work, we experimentally simulate a non-Hermitian, PT-symmetric AAH model using photonic quantum-walk dynamics. Building upon our previous experiments of quantum walks with non-Hermitian skin effects~\cite{photonskin,XDW+21}, we further introduce position- and polarization-dependent phase operators to implement quasiperiodicity. Through dynamic observables such as the Lyapunov exponents~\cite{stefano,xueexp}, the dynamic inverse participation ratio (dIPR)~\cite{crit3}, and the time-evolved photon counts, we confirm key properties of the phases on either side of the multi-critical point, as predicted by the spectral winding number. Interestingly, the measured dIPR further provides evidence for the presence of a mobility edge in the system, whereas mobility edges are known to be absent in the original AAH model.

\begin{figure*}[tbp]
\centering
\includegraphics[width=0.7\textwidth]{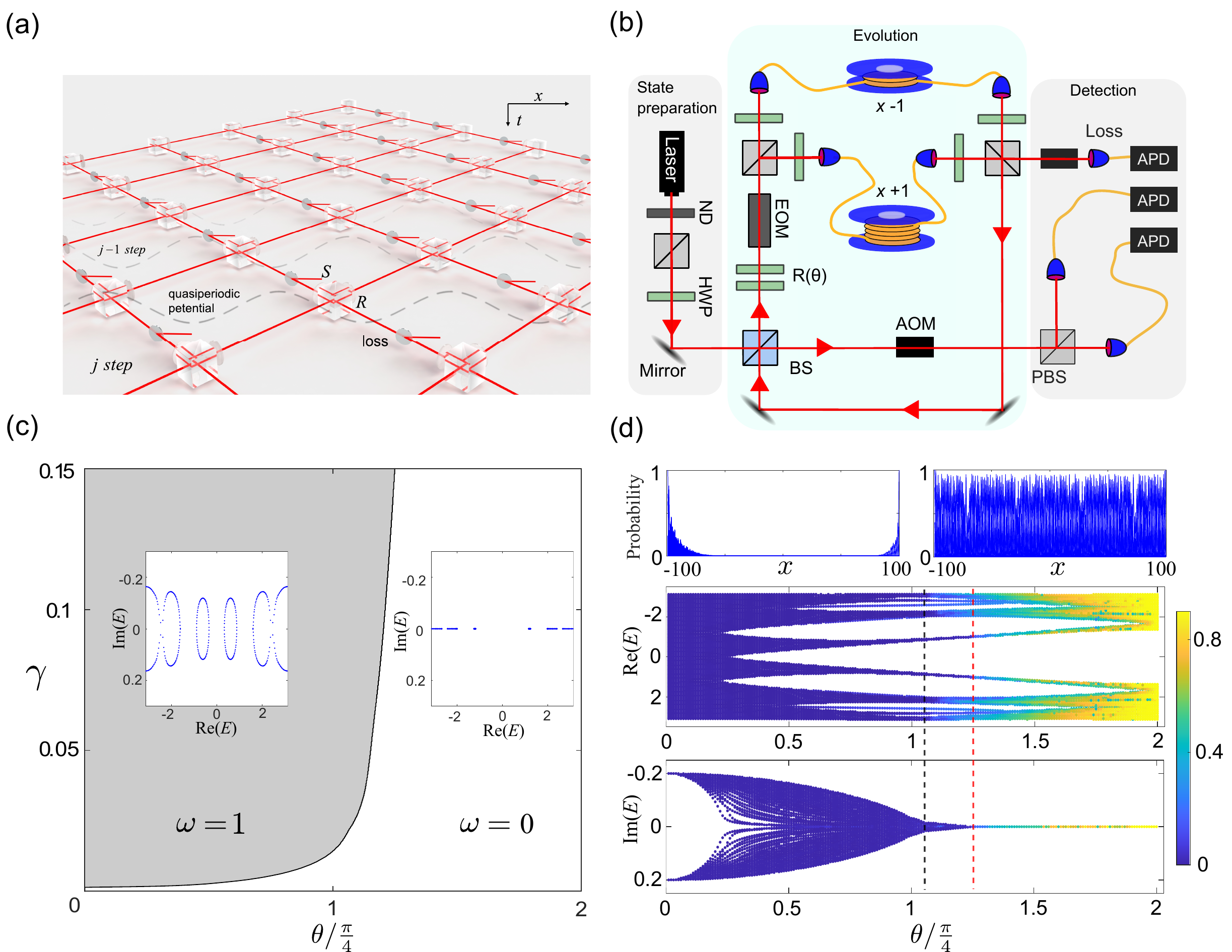}
\caption{Simulation of non-Hermitian quasicrystals. (a) A cartoon illustration of quantum-walk dynamics.
Operators $S$ and $R$ are defined in the main text. Lattice sites are located at the crossing points of the laser beams (red). The dashed grey curves illustrate the quasiperiodic phase (or a quasiperiodic potential for the effective Hamiltonian) introduced by the phase operator $P_{1,2}$.
(b) A time-multiplexed implementation of the photonic quantum walk illustrated in (a). ND: neutral density filter; HWP: half-wave plate; AOM: optical switch acousto-optic modulator; EOM: electro-optic modulator; (P)BS: (polarizing) beam splitter; APD: avalanche photo-diode. (c) Phase diagram of the spectral winding number $\omega$. Insets are the typical eigenspectra of the two phases on the complex plane with a fixed $\gamma=0.1$, and $\theta=\pi/8$ for $\omega=1$ and $\theta=3\pi/8$ for $\omega=0$. The phase boundary here not only marks the onset of the non-Hermitian skin effect, but also the PT-symmetry breaking point and the delocalization-localization transition. (d) Top panel: spatial distribution of all eigenstates of $U$ under the OBC, for the phase with $\omega=1$ (left) and $\omega=0$ (right), respectively. Middle panel:
The real components of the eigenergies $E$ of the effective Hamiltonian $H$ (under PBC) with increasing $\theta$, colored according to their respective inverse participation ratio $\text{IPR}$.
Lower panel: The imaginary components of the eigenenergies with increasing $\theta$ (under PBC), colored according to the $\text{IPR}$. For the numerical simulations, we take the lattice size $N=200$ and $\gamma=0.1$.
}
\label{fig:fig1}
\end{figure*}

\begin{figure}[tbp]
\includegraphics[width=0.5\textwidth]{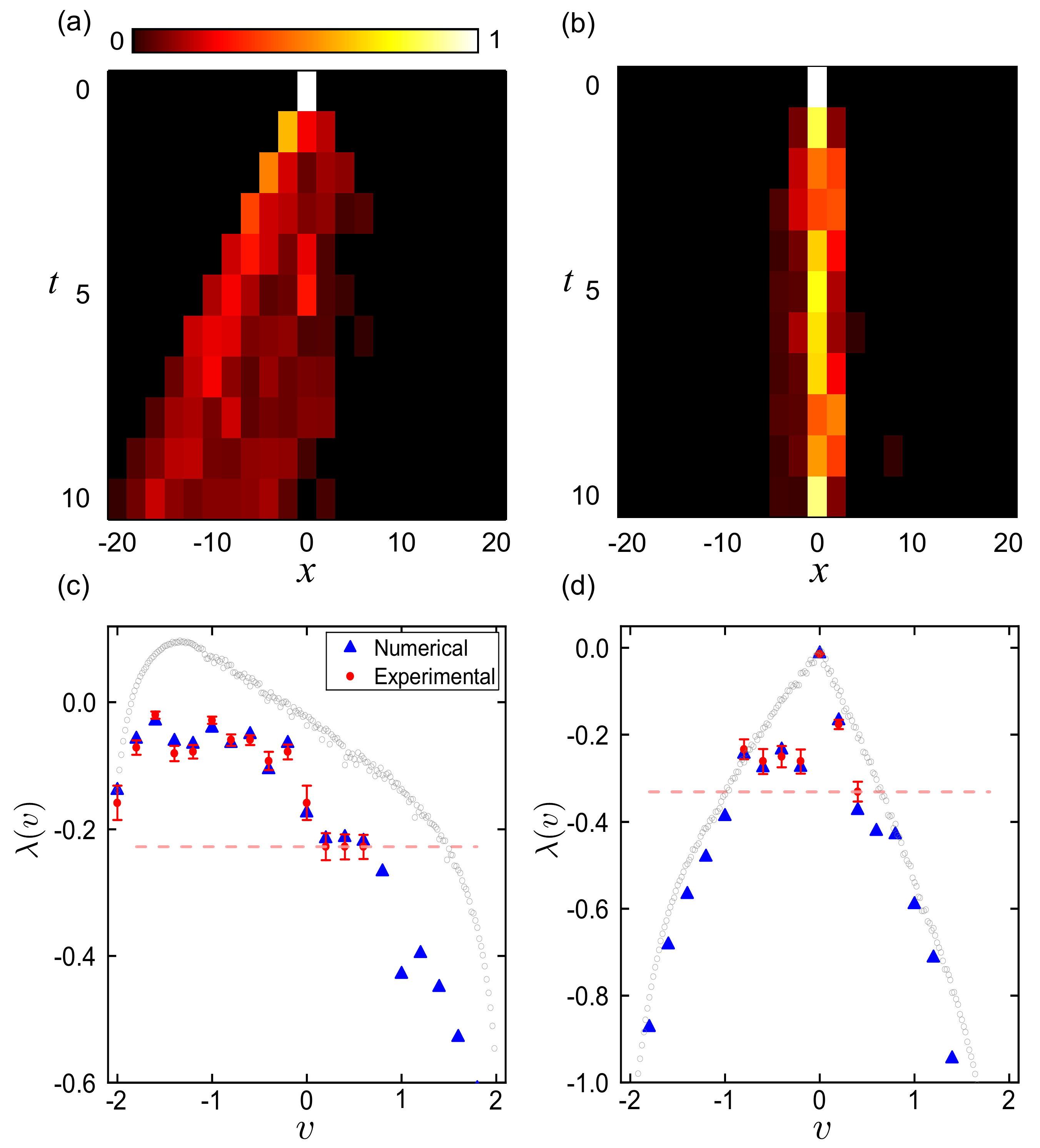}
\caption{Experimental detection of the non-Hermitian skin effect in a non-Hermitian quasicrystal. Measured photon distributions up to $10$ time steps with a fixed $\gamma=0.1$. The initial state is $\ket{x=0}\otimes\ket{H}$, and the coin parameters are $\theta=0.18\pi$ in (a) and $\theta=0.33\pi$ in (b), respectively. The corresponding numerical results are shown in the Supplemental Material. (c)(d) Measured Lyapunov exponent $\lambda(v)$ versus the shift velocity $v$ with the same parameters in (a) and (b). Red dots represent the experimental data, and blue triangles are the corresponding numerical simulations. Grey dotted lines denote the numerical results from $100$-step quantum walks. Orange dashed lines indicate the threshold values below which experimental data are no longer reliable due to photon loss. Error bars are due to the statistical uncertainty in photon-number counting.
}
\label{fig:fig2}
\end{figure}

{\it A time-multiplexed non-unitary quantum walk.---}
We simulate non-Hermitian quasicrystals using a one-dimensional photonic quantum walk, governed by the Floquet operator
\begin{align}
U=MSP_2R(\theta)MSP_1R(\theta),
\label{eq:U}
\end{align}
where the coin operator
$R(\theta)=\sum_x\ket{x}\bra{x}\otimes
\begin{pmatrix}
\cos\theta & -\sin\theta \\
\sin\theta & \cos\theta
\end{pmatrix}$,
under the polarization basis $\{|H\rangle,|V\rangle\}$, with $|H\rangle$ ($|V\rangle$) the horizontally (vertically) polarized state and $x$ labeling the lattice sites. The shift operator is given by
$S=\sum_x \ket{x-1}\bra{x}\otimes\ket{H}\bra{H}+\ket{x+1}\bra{x}\otimes\ket{V}\bra{V}$. Non-Hermiticity is introduced through the partial measurement $M=\sum_x\ket{x}\bra{x}\otimes
\begin{pmatrix}
e^{\gamma} & 0 \\
0 & e^{-\gamma}
\end{pmatrix}$ with $\gamma$ the gain-loss parameter.
The position-dependent phase operators satisfy $P_1=P_2^{-1}=\sum_x\ket{x}\bra{x}\otimes
\begin{pmatrix}
e^{\frac{i}{2}\cos(x\pi\phi)\pi} & 0 \\
0 & e^{-\frac{i}{2}\cos(x\pi\phi)\pi}
\end{pmatrix}$ with $\phi=(\sqrt{5}-1)/2$, the inverse of the golden Mean.

For the quantum-walk dynamics, the Floquet operator $U$ repeatedly acts on the initial state $|\Psi(0)\rangle$, with the time-evolved state given by $|\Psi(t)\rangle=U^t|\Psi(0)\rangle$, where $t$ labels the discrete time steps (see Fig.~\ref{fig:fig1}(a) for a cartoon illustration of the state evolution). The quantum walk is therefore a stroboscopic simulation of the time evolution driven by an effective Hamiltonian $H$, defined as $U=e^{-iH}$~\cite{photonskin,XDW+21}.

In the absence of the phase operators $P_i$ ($i=1,2$), the quantum walk governed by $U$ features non-Hermitian skin effect~\cite{photonskin,xueexp}. Therein, the interplay of the effective spin-orbit coupling (the coupling between polarization and spatial modes) and the polarization-dependent loss, leads to a finite, non-reciprocal probability current in the bulk that is responsible for the accumulation of population at boundaries, the namesake phenomenon of non-Hermitian skin effects. The quasiperiodicity is then introduced through the phase operators $P_{1,2}$, which can be understood as a polarization-dependent quasiperiodic potential for the effective Hamiltonian. The resulting Floquet operator $U$ thus realizes a stroboscopic simulation of the dynamics of a generalized non-Hermitian AAH model that features non-reciprocal hopping and quasiperiodic modulation of the on-site potential~\cite{supp}.

As illustrated in Fig.~\ref{fig:fig1}(b), we adopt a time-multiplexed scheme for the experimental implementation of $U$~\cite{supp}. Pulses from the photon source are attenuated to the single-photon level using a neutral density filter, ensuring a negligible probability of multi-photon events~\cite{sch,SGR+12}. Similar to the setup in Ref.~\cite{xueexp}, the various operators are implemented using beam splitters and half-wave plates, and are integrated into two optical fibre loops through which each photon passes twice for each discrete time step. Here different walker positions are encoded into the time domain, thanks to the polarization-dependent time delay introduced through the two optical fibre loops. With these, the position-dependent phase operators $P_{1,2}$ are realized through an electro-optical modulator that imposes a time-dependent phase within each discrete time step. Finally, we implement a polarization-dependent loss operator $M_\text{E}=e^{-\gamma}M$, and the experimentally realized time-evolved state $|\Psi_\text{E}(t)\rangle$ is related to
$|\Psi(t)\rangle$ through $|\Psi(t)\rangle=e^{2\gamma t}|\Psi_\text{E}(t)\rangle$.

\begin{figure}[tbp]
\includegraphics[width=0.5\textwidth]{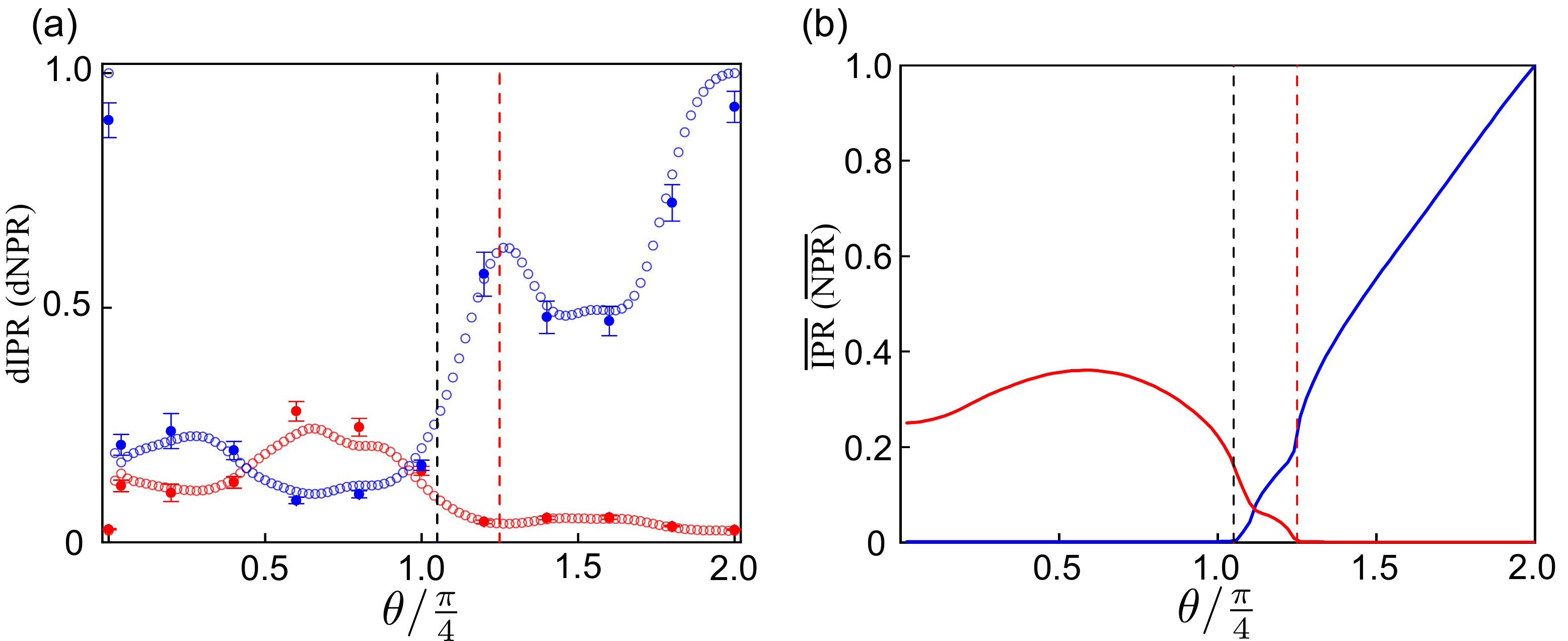}
\caption{Observation of the Anderson localization in a non-Hermitian quasicrystal. (a) Measured dIPR (blue) and dNPR (red) for non-Hermitian quantum walks with the initial state $\ket{x=0}\otimes\ket{V}$. Solid and hollow symbols represent the experimental data and their theoretical predictions, respectively. 
(b) Numerical simulation of $\overline{\text{IPR}}$ (blue) and $\overline{\text{NPR}}$ (red) for a lattice size $N=1000$. See the main text for the definitions of $\overline{\text{IPR}}$ and $\overline{\text{NPR}}$. Here we take $\gamma=0.1$.
}
\label{fig:fig3}
\end{figure}

{\it Winding number and phase diagram.---}
The quasicrystalline features of $U$ are best illustrated using the phase diagram characterized by a spectral winding number.
To calculate the winding number, we consider a periodic boundary condition (PBC) with $N$ lattices sites, and with an auxiliary magnetic flux $\Phi$ threaded through the resulting ring. This is equivalent to making the replacement $\gamma\rightarrow \gamma+i\Phi/(2N)$ in $U$. Subsequently, the spectral winding number is defined as~\cite{nonHtopo1,chen3}
\begin{align}
\omega=\lim_{N\rightarrow \infty}\frac{1}{2\pi i }\int^{2\pi}_{0}d\Phi\frac{\partial_\Phi \det\{H(\Phi/N)-E\}}{\det\{H(\Phi/N)-E\}},
\end{align}
where $H$ is the effective Hamiltonian of $U$, and $E$ is a base point in the complex energy plane. The quantum walk has a spectral topology, if there exists an $E$ such that $\omega\neq 0$. In this case, the PBC eigenspectrum of the effective Hamiltonian exhibits closed loops around the base point $E$, implying a persist current in the bulk which is the origin of the non-Hermitian skin effect under an OBC~\cite{fangchenskin,kawabataskin}. Otherwise, when $\omega= 0$, the PBC eigenspectrum of $H$ does not have spectral topology, and the quantum walk does not show the non-Hermitian skin effect under OBC. We show in Fig.~\ref{fig:fig1}(c), the phase diagram of the system. In the region with $\omega=1$, the emergence of non-Hermitian skin effect is confirmed in the top panel of Fig.~\ref{fig:fig1}(d), where all eigen wave functions accumulate toward the boundaries under OBC.

Under quasiperiodicity, eigenstates of the system undergo a delocalization-localization transition with increasing $\theta$.
The localization of the $n$th eigenstate can be characterized by the inverse participation ratio (IPR)~\cite{crit2} defined as
\begin{align}
\text{IPR}_n=\sum_x \frac{|\psi_n(x)|^4}{\sum_x |\psi_n(x)|^2},
\end{align}
where $\psi_n(x)$ is the support of the $n$th eigenstate of $U$ on site $x$. The state is delocalized (localized) when $\text{IPR}_n$ is vanishingly small (finite).
As illustrated in the middle panel of Fig.~\ref{fig:fig1}(d), the critical $\theta$ appears to be energy-dependent across the eigenspectrum, indicating the existence of a mobility edge.

Interestingly, the aforementioned spectral topological phase boundary simultaneously marks the transition point beyond which all eigenstates of the system are localized. In Fig.~\ref{fig:fig1}(d), we indicate such a global delocalization-location transition with a dashed vertical line in red. To the right of the line, all $\text{IPR}_n$ are finite, the system is fully localized, and the winding number $\omega=0$.
To the left of the line, not all eigenstates are localized (some states still feature finite $\text{IPR}_n$), and the winding number $\omega=1$. One may further identify a transition point [indicated by the black dashed line
in Fig.~\ref{fig:fig1}(d)] where the first localized eigenstates (the ground and the highest excited states) emerge in the spectrum.
This transition marks the onset of the mobility edge as $\theta$ increases, but does not coincide with the spectral topological transition.

%

%
%

The spectral topological phase boundary is also the critical point at which the PT-symmetry becomes broken under PBC. Due to the spectral topological origin of the non-Hermitian skin effect, the region with $\omega=1$ is necessarily PT-symmetry broken. Intriguingly, as shown in the lower panel of Fig.~\ref{fig:fig1}(d), in the region with $\omega=0$, the eigensepctrum of $H$ is entirely real, indicating the concurrence of the spectral topological transition, the global delocalization-localization transition, and the PT-symmetry breaking transition~\cite{chen3,supp}.
In the following, we experimentally probe these transitions one-by-one.

{\it Non-Hermitian skin effects.---}
We experimentally probe the presence of the non-Hermitian skin effects through bulk dynamics~\cite{xueexp}.
While the direct manifestation of the non-Hermitian skin effect is the accumulation of population at open boundaries, it originates from a persistent, directional bulk current that can be directly probed through bulk dynamics. For this purpose, we initialize the walker near $x=0$, and record the photon distribution at each time step up to $t=10$. For $\theta<\theta_c$, the system possesses non-Hermitian skin effects. This is reflected as the directional probability flow in the bulk, as shown in Fig.~\ref{fig:fig2}(a). By contrast, for sufficiently large $\theta$, the system is in the Anderson-localized state, and the probability would accumulate at the initial position, as shown in Fig.~\ref{fig:fig2}(b).

The conclusion, while consistent with the phase diagram, is further confirmed by measuring the Lyapunov exponent, defined as~\cite{stefano,xueexp}
\begin{align}
\lambda(v)=\frac{1}{t}\log|\left(\langle x=vt|\otimes\langle H|\right)|\Psi(t)\rangle|,
\end{align}
where $v$ is the shift velocity. The Lyapunov exponent reveals the presence of the bulk probability flow, since by definition, the location of its peak shows how the wave function propagates along the lattice.
Specifically, $\lambda(v)$ peaks at a finite shift velocity $v$ in the presence of non-Hermitian skin effect; whereas it peaks at $v=0$ in the absence of non-Hermitian skin effect~\cite{stefano,xueexp}.
Experimentally, we construct $\lambda(v)$ following a $10$-step quantum walk, by measuring the photon distribution at each step. As shown in Figs.~\ref{fig:fig2}(c)(d), the experimental data confirms the presence of non-Hermitian skin effect in the region with $\omega=1$.

\begin{figure}[tbp]
\centering
\includegraphics[width=0.5\textwidth]{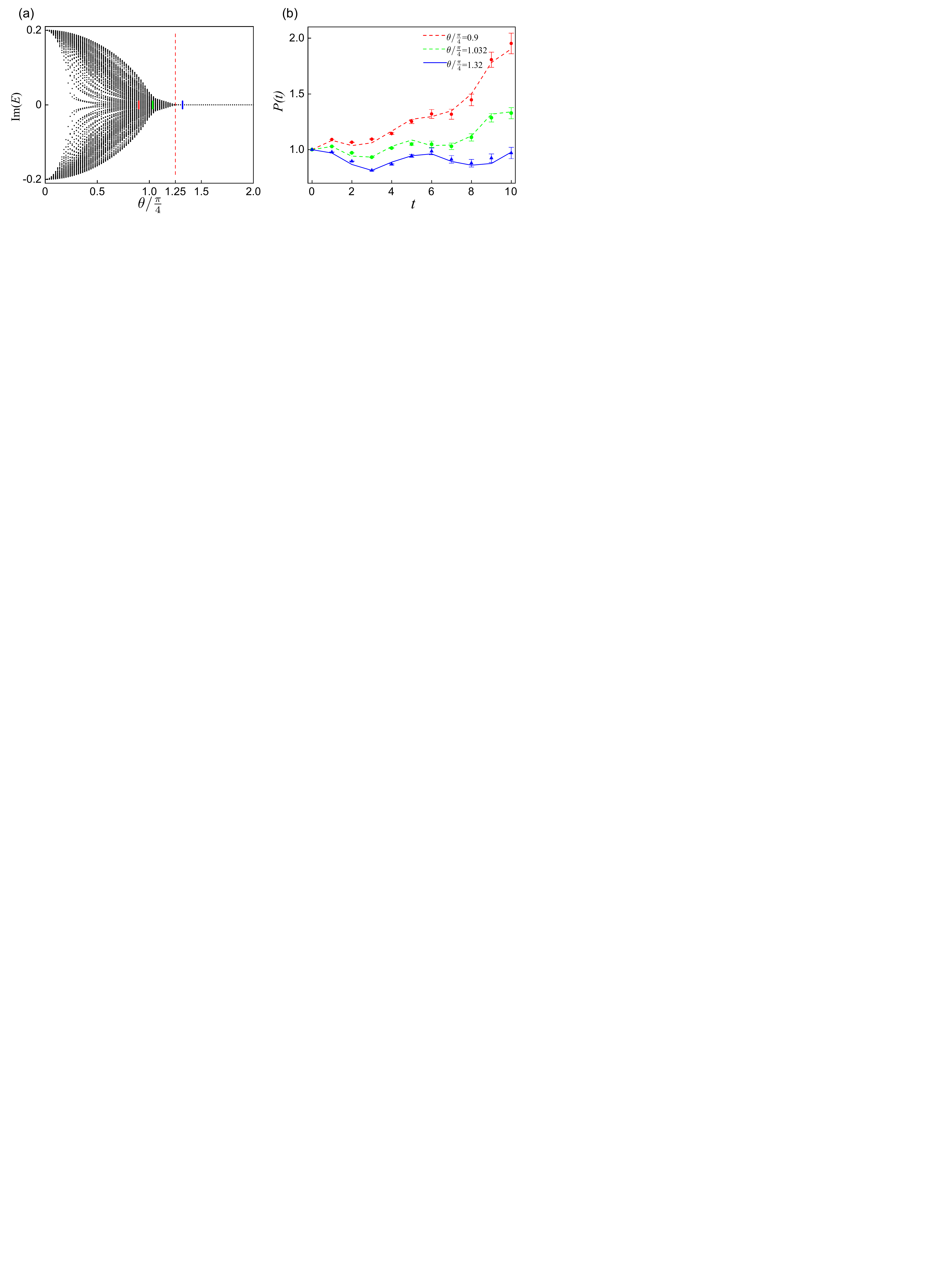}
\caption{PT-symmetry transition in a non-Hermitian quasicrystal. (a) Imaginary parts of the quasienegies of a non-Hermitian quantum walk with a lattice size $N=200$, and $\gamma=0.1$. The vertical dashed line at $\theta/\frac{\pi}{4}\approx 1.25$ separate the PT unbroken and broken phases. (b) Measured overall corrected probabilities $P(t)$ of the quantum walk with an initial state $\ket{x=0}\otimes\ket{H}$ for various $\theta$ corresponding to the PT unbroken and broken phases. Symbols represent the experimental data and curves for their theoretical predictions. 
}
\label{fig:fig4}
\end{figure}

{\it Localization and mobility edge.---}
To characterize the localization properties, we measure the dIPR~\cite{crit3}, defined as $\text{dIPR}=\sum_{x} P^2(x,t)$, where $P(x,t)$ is the corrected probability of the walker on position $x$ at the end of time step $t$.
The probability is normalized by dividing the total photon count after the $t$-th step, so that $P(x,t)= |\langle x|\Psi_\text{E}(t)\rangle|^2/\sum_x |\langle x|\Psi_\text{E}(t)\rangle|^2$.

A vanishingly small dIPR at long times suggests the delocalized, metallic phase; while a finite dIPR indicates the onset of localization.
As illustrated in Fig.~\ref{fig:fig3}(a), the measured dIPR (blue) indicates a transition point near $\theta/\frac{\pi}{4}\approx 1.05$, which is consistent with the black vertical dashed line in Fig.~\ref{fig:fig1}(d). This is the location where some eigenstates start to be localized, as a mobility edge emerges in the eigenspectrum.

To reveal the global delocalization-localization transition where all eigenstates become Anderson localized, we construct the dynamic normalized inverse participation ratio (dNPR), defined as $\text{dNPR}=[2t\sum_{x} P^2(x,t)]^{-1}$. As shown in Fig.~\ref{fig:fig3}(a), the measured dNPR (red) becomes vanishingly small near $\theta/\frac{\pi}{4}\approx 1.25$, consistent with the global delocalization-localization transition in Fig.~\ref{fig:fig1}(d).

The measured dIPR and dNPR are respectively the dynamic probe to the averaged IPR: $\overline{\text{IPR}}=(1/2N)\sum_n \text{IPR}_n$; and the average normalized IPR: $\overline{\text{NPR}}=1/(4N^2)\sum_n 1/\text{IPR}_n$~\cite{huse,chen4}. Key features in the numerical simulations of these quantities agree well with our measurements [see Fig.~\ref{fig:fig3}(b)], indicating a mobility edge in the range $\theta/\frac{\pi}{4}\in (1.05,1.25)$.


%
%

{\it Breaking parity-time symmetry.---}
We confirm the breaking of PT symmetry, by measuring the time evolution of the overall corrected probability of photons. Here the overall corrected probability is defined as $P(t)=e^{4\gamma t}\sum_x |\langle x|\Psi_E(t)\rangle|^2$, where the exponent $e^{4\gamma t}$ recovers the probability evolution under $U$. When the system is in the PT unbroken regime, the quasienergy is entirely real, and $P(t)$ would be on the order of unity. By contrast, when the system is in the PT broken regime, $P(t)$ should grow in time, as the eigenenergies can acquire positive imaginary components. These understandings are confirmed in Fig.~\ref{fig:fig4}, where time evolutions at different $\theta$ shows that the PT transition point is consistent with the theoretical phase diagram.
%

{\it Conclusion.---}
We experimentally simulate the dynamics of a one-dimensional non-Hermitian quasicrystal using photonic quantum walks, and reveal an
tricritical point where the spectral topological transition, the global delocalization-localization transition, and the PT-symmetry breaking transition simultaneously occur. Our experiment thus unveils a highly non-trivial phenomenon that is absent in Hermitian quasicrystals, and calls for further study of the critical phenomena near these phase transitions.

Intriguingly, the coincidence of all three phase transitions can be model-dependent. In a very recent theoretical study~\cite{naad}, it is shown that, although the localization and the PT-symmetry breaking transitions coincide in a general class of non-Hermitian quasicrystals, they are not necessarily accompanied by a change of the spectral topology. While the study raises important open questions regarding the conditions for the concurrence of phase transitions, our experiment offers a highly tunable platform on which these questions can be systematically addressed in the future.

\begin{widetext}
\renewcommand{\thesection}{\Alph{section}}
\renewcommand{\thefigure}{S\arabic{figure}}
\renewcommand{\thetable}{S\Roman{table}}
\setcounter{figure}{0}
\renewcommand{\theequation}{S\arabic{equation}}
\setcounter{equation}{0}

\section{Supplemental Material for ``Simulating non-Hermitian quasicrystals with single-photon quantum walks''}

\subsection{Experimental scheme}

We adopt a time-multiplexed scheme for the experimental simulations of non-Hermitian quasicrystals with single-photon quantum walks~\cite{sch,SGR+12,scienceskin}. The photon source is provided by a pulsed laser with a central wavelength of $808$nm, a pulse width of $88$ps, and a repetition rate of $31.25$KHz. The pulses are attenuated to the single-photon level by using a neutral density filter. An average photon number per pulse is less than $2\times 10^{-4}$, which ensures a negligible probability of multi-photon events. Photons pass through a polarizing beam splitter (PBS) and a half-wave plate (HWP) for initializing the polarization states. In this process, the coins of the quantum walk are encoded in the polarization state, and are coupled in and out of a time-multiplexed setup through a beam splitter (BS) with a reflectivity of $5\%$, corresponding to a low coupling rate of photons into the network. Such a low-reflectivity BS also enables the out-coupling of photons for measurement.

The setup consists of two optical fibre loops with different lengths, which are coupled by PBSs. The shift operator $S$ is implemented by separating photons corresponding to their two polarization components and routing them through the two fibre loops, respectively. Polarization-dependent time delay is then introduced and the walker's position is encoded into the time domain. The lengths of the two fiber loops are $167.034$m and $160.000$m, respectively. A round-trip time is approximately given by $772$ns. The time difference of photons traveling through two fiber loops is $33.046$ns, which defines the temporal width of a time-bin. To implement the coin operator $R(\theta)$, two HWPs provide a careful control over the parameter $\theta$.
The position-dependent phase operator is implemented by an electro-optical modulator (EOM)
\begin{equation}
P=\sum_x\ket{x}\bra{x}\otimes
\begin{pmatrix}
e^{i\frac{\phi'}{2}} & 0 \\
0 & e^{-i\frac{\phi'}{2}}\end{pmatrix}
\end{equation}
with $\phi'=\cos(x\pi\phi)\pi$. The rise/fall times of EOM ($4$ns) are much less than the time difference of the adjacent position ($33.046$ns), which enables us to control the parameter $\phi^{\prime}(x)$ precisely.

To realize a polarization-dependent loss operation $M_\text{E}=e^{-\gamma}M$, two HWPs are introduced into each of the fiber loops. For the short loop, after passing through the first PBS, horizontally polarized photons are all transmitted by the second PBS and are subject to further time evolution. Whereas for the long loop, by controlling the setting angle of the HWP, part of the photons $(1-e^{4\gamma})$ with vertical polarization are flipped into horizontal ones, transmitted by the second PBS and subsequently leak out of the setup. Then we implement a half step of the split-step quantum walk $U$. We therefore read out the time-evolved state driven by $U$ by adding a time-dependent factor $e^{2\gamma t}$ to the outcomes of our experimental measurement. For instance, a superposition of multiple spatial positions at a given time step is translated into the superposition of multiple well-resolved pulses within the same discrete time step.

Three avalanche photo-diodes (APDs) are employed to record the temporal and polarization properties of both the out-coupled photons and the lost photons, yielding information regarding the number of time steps, as well as the spatial and coin states of the walker.

The probability distribution of the walker $P(x,t)=\frac{|\langle x|\Psi_{\text{E}}(t)\rangle|^2}{\sum_x |\langle x|\Psi_{\text{E}}(t)\rangle|^2}$ at the position $x$ and step $t$ is obtained by dividing the number of photons collected by APD at the position $x$ using the total photons collected after the $t$th step, i.e.,
\begin{equation}
P(x,t)=\frac{N(x,t)}{\sum_x N(x,t)}
\end{equation}
with $N(x,t)=N_H(x,t)+N_V(x,t)$. With the probability distribution $P(x,t)$, we can construct the dynamic IPR $\text{dIPR}=\sum_x P^2(x,t)$.
Note that, as defined in the main text, $|\Psi(t)\rangle=e^{2\gamma t}|\Psi_E(t)\rangle$, where $|\Psi(t)\rangle$ and $|\Psi_\text{E}(t)\rangle$ are respectively the time-evolved states under $U$, and under the experimentally implemented dynamics.

The overall probability $P(t)=\sum_x|\langle x|\Psi(t)\rangle|^2$ can be constructed as
\begin{equation}
P(t)=e^{4\gamma t}\prod_{t'=1}^t\left[\frac{\frac{N'(t')}{r}}{\frac{N'(t')}{r}+N_{1L}(t')}\times\frac{\frac{N(t')}{r}}{\frac{N(t')}{r}+N_{2L}(t')}\right].
\end{equation}
The photon loss caused by the partial measurement $M$ is also included. Here $r=0.05$ is the reflectivity of the BS, $N'(t)=\sum_x N'(x,t)$ is the total photon numbers reflected by the BS after the first half step of the $t$th step, $N(t)=\sum_x N(x,t)$ is the total photon numbers reflected by the BS after the $t$th step, and $N_{1L,2L}(t)=\sum_x N_{1L,2L}(x,t)$ are the total photon loss after the first (second) half step of the $t$th step. Then, we can construct the Lyapunov exponent
\begin{equation}
\lambda(v)=\frac{1}{t}\log\sqrt{\frac{P(t)N_H(x,t)}{\sum_x N_H(x,t)+\sum_x N_V(x,t)}}
\end{equation}
with the experimental results of $P(t)$, where $N_{H,V}(x,t)$ are the photon numbers with horizontal (vertical) polarizations collected by APDs at the position $x$ after the $t$th step.

\subsection{Dynamics of the time-evolved state}

The quantum-walk dynamics implemented in our experiment constitutes a stroboscopic simulation of the time evolution driven by a generalized non-Hermitian Aubry-Andr\'e-Harper (AAH) model with both diagonal and off-diagonal quasiperiodic disorder.

To see this, we consider the state at the $m$-th time step $|\Psi\rangle=\sum_{x}\left(\begin{array}{l}\mu^m_{x} \\ v^m_{x}\end{array}\right) \otimes|x\rangle$.
Applying $U$ one on the state once, we obtain the dynamic equation, connecting the state at the $m+1$-th step with that at the $m$-th step
\begin{align}
\mu_{x}^{m+1}=&\left(\cos^2 (\theta) \mu_{x+2}^{m}-\cos(\theta)\sin (\theta) v_{x+2}^{m}\right) \exp \left(2\gamma+\frac{i \phi^{\prime}(x+2)-i\phi^{\prime}(x+1)}{2}\right)\nonumber\\
&-\left(\sin^2(\theta)\mu_{x}^{m}+\cos(\theta)\sin (\theta) v_{x}^{m}\right) \exp \left(\frac{i \phi^{\prime}(x)+i\phi^{\prime}(x+1)}{2}\right),\label{eq:mux}\\
v_{x}^{m+1}=&\left(\cos (\theta)\sin(\theta) \mu_{x-2}^{m}+\cos^2(\theta)v_{x-2}^{m}\right) \exp \left(-2\gamma-\frac{i \phi^{\prime}(x-2)-i\phi^{\prime}(x-1)}{2}\right)\nonumber\\
&+\left(\cos (\theta)\sin(\theta)\mu_{x}^{m}-\sin^2 (\theta) v_{x}^{m}\right) \exp \left(\frac{i \phi^{\prime}(x)+i\phi^{\prime}(x-1)}{2}\right),\label{eq:vx}
\end{align}
where $\phi^{\prime}(x)=\cos (x \pi \phi) \pi$.

From the above two equations, we have $\sum_{x}\left(\begin{array}{c}\mu_{x}^{m+1} \\ \boldsymbol{v}_{x}^{m+1}\end{array}\right)|x\rangle=U \sum_{x}\left(\begin{array}{c}\mu_{x}^{m} \\ \boldsymbol{v}_{x}^{m}\end{array}\right)|x\rangle$, where
\begin{align}
U_{11}&=\sum_{x}-\sin ^{2}(\theta) \exp \left(\frac{i \phi^{\prime}(x)+i\phi^{\prime}(x+1)}{2}\right)|x\rangle\langle x|+\cos ^{2}(\theta)  \exp \left(2\gamma+\frac{i\phi^{\prime}(x+2)-i\phi^{\prime}(x+1)}{2}\right)| x\rangle\langle x+2|, \label{eq:U11}\\
U_{12}&=\sum_{x}-\sin (\theta) \cos (\theta) \exp \left(\frac{i \phi^{\prime}(x)+i \phi^{\prime}(x+1)}{2}\right)|x\rangle\langle x|-\sin (\theta) \cos (\theta) \exp \left(2\gamma+\frac{i \phi^{\prime}(x+2)-i \phi^{\prime}(x+1)}{2}\right)| x\rangle\langle x+2|, \label{eq:U12}\\
U_{21}&=\sum_{x} \sin (\theta) \cos (\theta) \exp \left(\frac{i \phi^{\prime}(x-1)+i \phi^{\prime}(x)}{2}\right)|x\rangle\langle x|+\sin (\theta) \cos (\theta) \exp \left(-2\gamma-\frac{i \phi^{\prime}(x-2)-i\phi^{\prime}(x-1)}{2}\right)| x\rangle\langle x-2|, \label{eq:U21}\\
U_{22}&=\sum_{x}-\sin ^{2}(\theta) \exp \left(\frac{i \phi^{\prime}(x-1)+i \phi^{\prime}(x)}{2}\right)|x\rangle\langle x|+\cos ^{2}(\theta) \exp \left(-2\gamma-\frac{i \phi^{\prime}(x-2)-i\phi^{\prime}(x-1)}{2}\right)| x\rangle\langle x-2|.\label{eq:U22}
\end{align}

For comparison, we derive the corresponding dynamic equation for the original AAH model below. We consider the Hamiltonian
\begin{align}
H=\sum_xJ(|x\rangle\langle x+1|+|x+1\rangle\langle x|)+\sum_xV(x)|x\rangle\langle x|,
\end{align}
where $J$ is the hopping rate, and $V(x)=V\cos(2\pi\phi x)$ is the on-site quasiperiodic potential, with $\phi$ an irrational number.

We define the infinitesimal time-evolution operator by $U_H=e^{-iH\Delta t}$, where the time interval $\Delta t$ is infinitesimal. Using the Baker–Campbell–Hausdorff formula, we have
\begin{align}
U_H=&e^{-iH\Delta t}=e^{-i\Delta tJ\sum_x(|x\rangle\langle x+1|+|x+1\rangle\langle x|)}e^{-i\Delta t\sum_xV(x)|x\rangle\langle x|}+o(\Delta t^2)\nonumber\\
=&\sum_x e^{-i\Delta t V(x)}|x\rangle\langle x|-i\Delta tJ\sum_x\left(e^{-i\Delta tV(x+1)}|x\rangle\langle x+1|+e^{-i\Delta tV(x)}|x+1\rangle\langle x| \right)+o(\Delta t^2).\label{eq:UH}
\end{align}

Now we consider the state $|\phi\rangle=\sum_{x}a^m_x|x\rangle$, where the superscript indicates the $m$-th time step, meaning the operator $U_H$  has been applied $m$ times on an initial state.
Applying $U_H$ one more time, and keeping terms of the order $\Delta t$, we have
\begin{align}
a^{m+1}_x= e^{-i\Delta t V(x)}a_x^m-i\Delta tJe^{-i\Delta tV(x+1)}a_{x+1}^m-i\Delta tJe^{-i\Delta tV(x-1)}a_{x-1}^m.\label{eq:amx}
\end{align}
Comparing Eqs.~(\ref{eq:mux})(\ref{eq:vx}) and Eq.~(\ref{eq:amx}) [or equivalently Eqs.~(\ref{eq:U11})(\ref{eq:U12})(\ref{eq:U21})(\ref{eq:U22}) and Eq.~(\ref{eq:UH})],
it is straightforward to see that the implemented quantum-walk dynamics corresponds to a discrete-time evolution governed by a lattice model with quasiperiodic potential (given by $\phi^{\prime}(x)$). Further, different signs of $\gamma$ in the exponentials of Eqs.~(\ref{eq:mux})(\ref{eq:vx}) indicate the lattice model features a non-reciprocal hopping, which eventually gives rise to the non-Hermitian skin effect.

\subsection{Reality of eigenenergy and localization}

From numerical simulations, it is straightforward to establish that for eigenstates with real eigenenergies, the state is spatially localized. For those with a finite imaginary components, the state is spatially extended. This is illustrated in Fig.~\ref{fig:figSpt}.

\begin{figure}[h]
\centering
\includegraphics[width=1\textwidth]{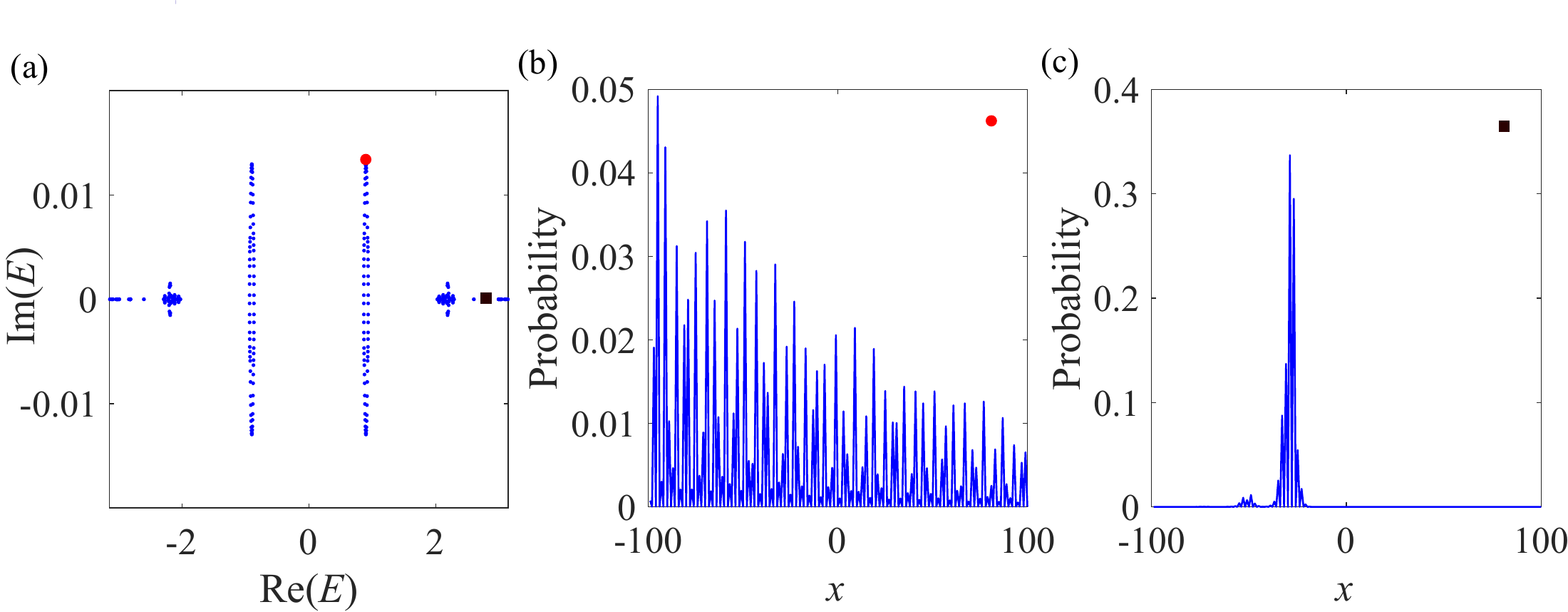}
\caption{(a) Eigenspectrum of the effective Hamiltonian $H$ in the PT-broken regime under a periodic boundary condition, where we take a lattice size $N=200$, $\gamma=0.1$, and $\theta/(\frac{\pi}{4})=1.1$ for numerical calculations. (b)(c) shows the spatial probability distribution of the eigenstates with eigenenergies marked by red dot (b) and black square (c) in the eigenspectrum (a), respectively.
}
\label{fig:figSpt}
\end{figure}

In the parity-time (PT) symmetry broken regime, some of the eigenstates can still have real eigenenergies; while in the PT symmetry unbroken regime, all eigenstates have real eigenenergies.
Thus, when the system crosses the PT symmetry breaking point from the PT-broken side to the unbroken side, a global delocalization-localization transition occurs, where all eigenstates are localized on the unbroken side. This explains the concurrence of the PT-symmetry breaking transition and the delocalization-localization transition at $\theta\approx 1.25\frac{\pi}{4}$.

\subsection{Dependence of the phase boundary on the gain-loss parameter}

As shown in the phase diagram in Fig.~\ref{fig:fig1} of the main text, the phase boundary is dependent on the gain-loss parameter $\gamma$. We experimentally confirm this by probing the PT-symmetry breaking and the delocalization-localization transitions by adopting a different $\gamma$ compared to the main text.

In Fig.~\ref{fig:figS3}(a), we show the numerically calculated $\overline{\text{IPR}}$ (blue) and $\overline{\text{NPR}}$ (red) for $\gamma=0.02$ and $N=1000$. The two transitions are now located at $\theta/\frac{\pi}{4}\approx 0.98$ and $\theta/\frac{\pi}{4}\approx 1.2$, respectively. The latter should correspond to the triple phase transition point. The experimentally measured dIPR and dNPR in Fig.~\ref{fig:figS3}(b) are consistent with the numerical simulations, indicating a global delocalization-localization transition near $\theta/\frac{\pi}{4}\approx 1.2$.

In Fig.~\ref{fig:figS3}(c), we show the numerically evaluated $\text{Im}(E)$ for $\gamma=0.02$ and $N=200$, where a PT-symmetry breaking transition is identified at $\theta/\frac{\pi}{4}\approx 1.2$. The experimentally measured overall corrected probability $P(t)$ for different $\theta$ in Fig.~\ref{fig:figS3}(d) are mostly consistent with the PT-symmetry breaking transition. For the odd $\theta/\frac{\pi}{4}=1.032$ (green), $P(t)$ should grow in time (see numerical simulation in the inset). Due to the limitation of experimentally accessible time steps and its closeness to the transition point, the measured $P(t)$ for $\theta/\frac{\pi}{4}=1.032$ appears to be on the order of unity. Nevertheless, together with the measured $P(t)$ for $\theta/\frac{\pi}{4}=1.032$ in Fig.~\ref{fig:fig4}(b) of the main text, our data provide evidence for the dependence of the PT transition on $\gamma$.

\begin{figure}[tbp]
\centering
\includegraphics[width=0.7\textwidth]{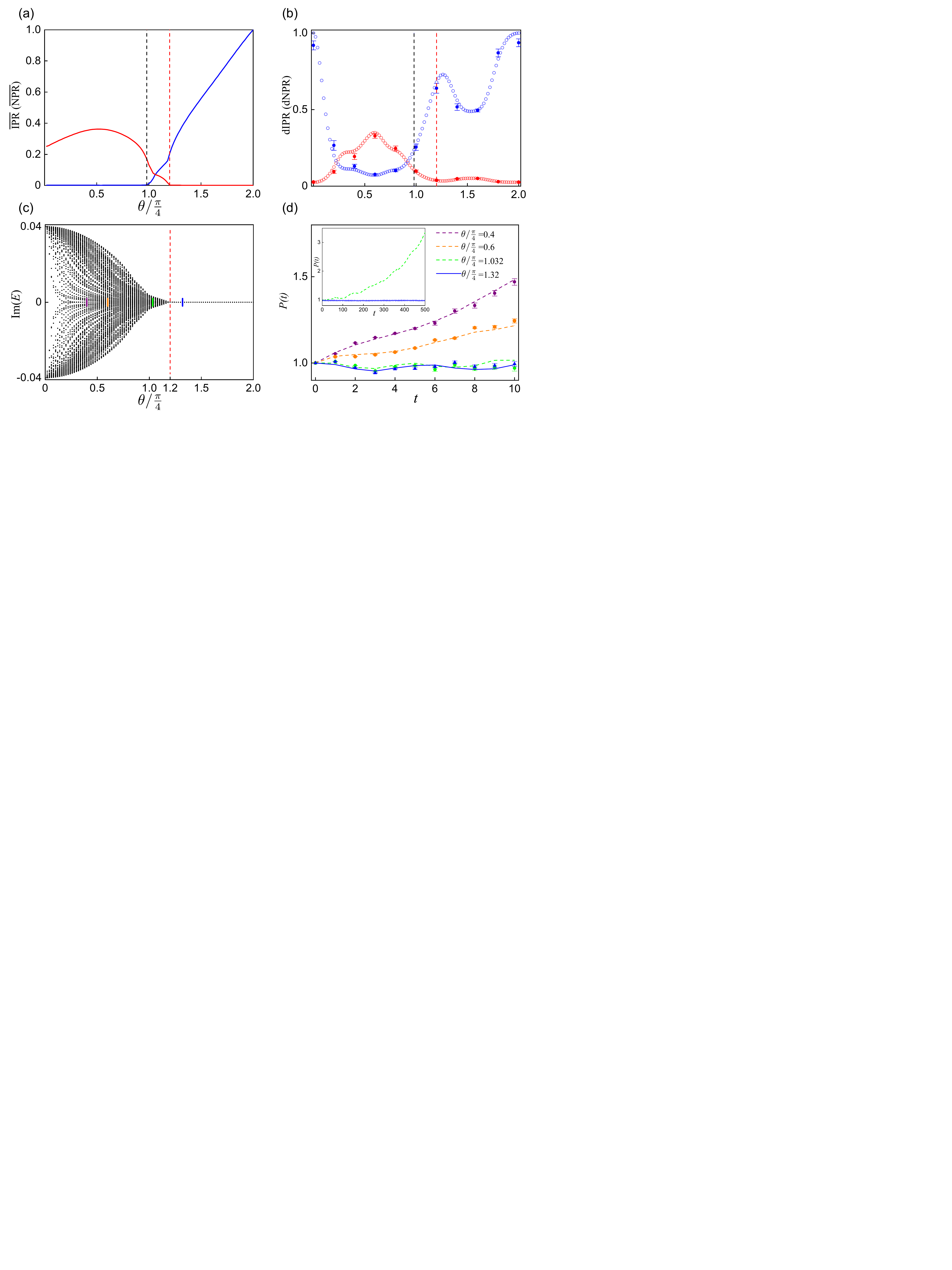}
\caption{(a) Numerical simulation of $\overline{\text{IPR}}$ (blue) and $\overline{\text{NPR}}$ (red) for
$N=1000$ and $\gamma=0.02$.
(b) Measured dIPR (blue) and dNPR (red) for non-Hermitian quantum walks with the initial state $\ket{x=0}\otimes\ket{V}$ and $\gamma=0.02$. Solid and hollow symbols represent the experimental data and their theoretical predictions, respectively. The black and red dashed lines in (a)(b) are at $\theta/\frac{\pi}{4}\approx 0.98$ and $\theta/\frac{\pi}{4}\approx 1.2$, respectively.
(c) Imaginary parts of the quasienegies of a non-Hermitian quantum walk with $N=200$ and $\gamma=0.02$. The vertical dashed line at $\theta/\frac{\pi}{4}\approx 1.2$ separate the PT unbroken and broken phases.
(d) Measured overall corrected probabilities $P(t)$ of a non-Hermitian quantum walk with the initial state $\ket{x=0}\otimes\ket{H}$ for various $\theta$ corresponding to the PT unbroken and broken phases. Symbols represent the experimental data and curves for their theoretical predictions. Error bars are due to the statistical uncertainty in photon-number counting. Inset: numerically simulated $P(t)$ at longer time steps for $\theta/\frac{\pi}{4}=1.32$ (blue) and $\theta/\frac{\pi}{4}=1.032$ (green), respectively.
}
\label{fig:figS3}
\end{figure}

\subsection{Additional numerical results}

Additional numerical results corresponding to Figs.~\ref{fig:fig2}(a)(b) in the main text are shown in Fig.~\ref{fig:figS1}.

\begin{figure}[h!]
\centering
\includegraphics[width=0.5\textwidth]{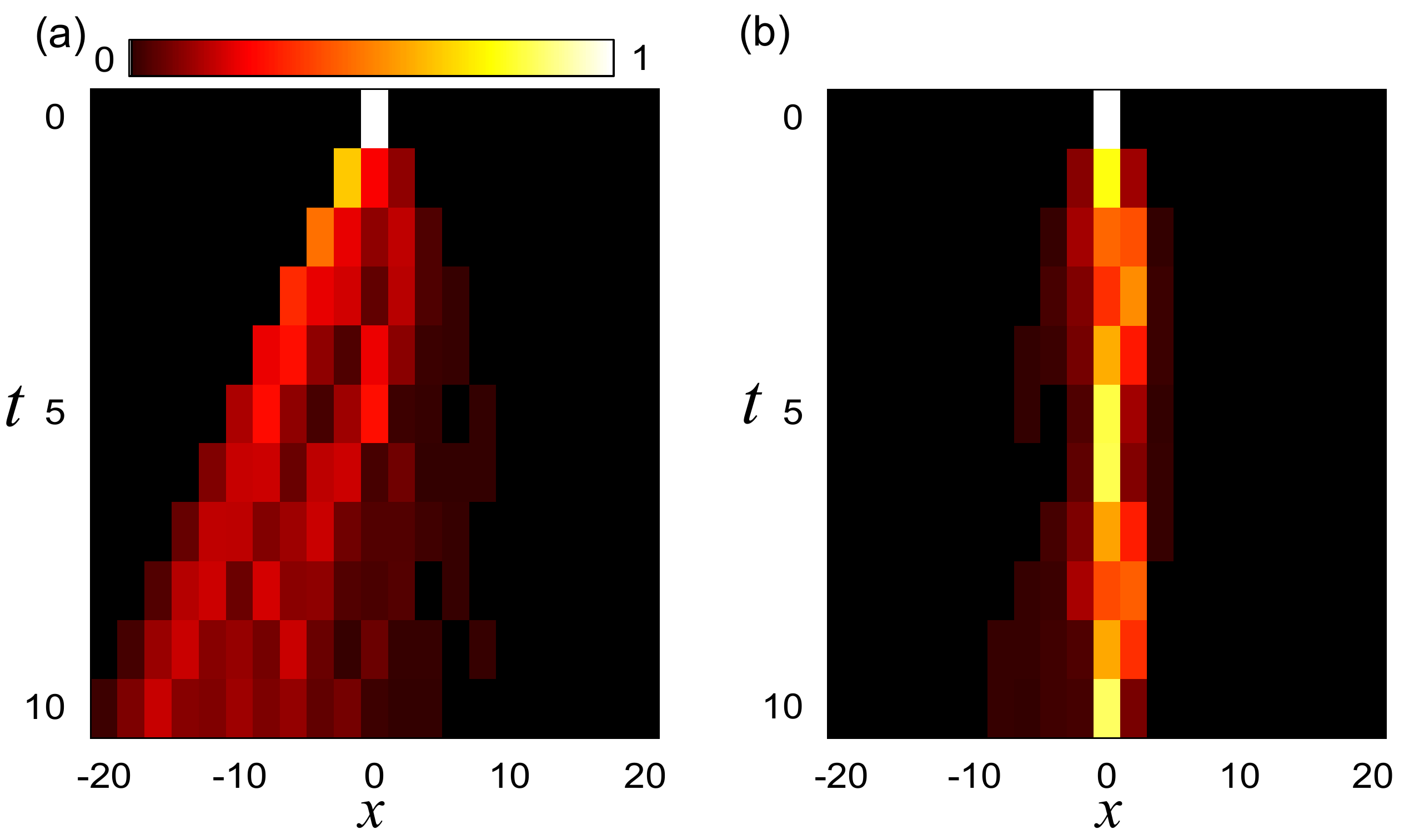}
\caption{The numerical simulations corresponding to experimental data in Figs.~\ref{fig:fig2}(a) and (b), respectively.
}
\label{fig:figS1}
\end{figure}

\end{widetext}

\begin{acknowledgments}
This work has been supported by the National Natural Science Foundation of China (Grant Nos. 12025401, U1930402, 11974331 and 12088101). W. Y. acknowledges support from the National Key Research and Development Program of China (Grant Nos. 2016YFA0301700 and 2017YFA0304100).
\end{acknowledgments}

\end{document}